 \documentclass[acmlarge, authorversion]{acmart}

\AtBeginDocument{%
  \providecommand\BibTeX{{%
    \normalfont B\kern-0.5em{\scshape i\kern-0.25em b}\kern-0.8em\TeX}}}

\setcopyright{acmlicensed}
\acmJournal{JOCCH}
\acmYear{2020} \acmVolume{0} \acmNumber{0} \acmArticle{0} \acmMonth{0} \acmPrice{15}
\acmDOI{10.1145/3439862}

\begin{document}

\title{We Dare You: A Lifecycle Study of a Substitutional Reality Installation in a Museum Space}


\author{Petros Ioannidis}
\email{peio@itu.dk}
\orcid{0000-0002-7090-6954}
\affiliation{%
  \institution{Center for Computer Games Research, IT University of Copenhagen}
  \streetaddress{Rued Langgaards Vej 7 }
  \city{København}
  \postcode{2300, Denmark}
}

\author{Lina Eklund}
\email{lina.eklund@im.uu.se}
\orcid{0000-0002-8086-4962}
\affiliation{%
  \institution{Department of Informatics and Media, Uppsala University}
  \streetaddress{Box 513}
  \city{Uppsala}
  \postcode{751 20, Sweden}
}

\author{Anders Sundnes Løvlie}
\email{asun@itu.dk}
\orcid{0000-0003-0484-4668}
\affiliation{%
  \institution{Center for Computer Games Research, IT University of Copenhagen}
  \streetaddress{Rued Langgaards Vej 7 }
  \city{København}
  \postcode{2300, Denmark}
}


\begin{abstract}
In this article, we present a lifecycle study of \textit{We Dare You}, a Substitutional Reality (SR) installation that combines visual and tactile stimuli. The installation is set up in a center for architecture, and invites visitors to explore its facade while playing with vertigo, in a visual Virtual Reality (VR) environment that replicates the surrounding physical space of the installation. Drawing on an ethnographic approach, including observations and interviews, we researched the exhibit from its opening, through the initial months plagued by technical problems, its subsequent success as a social and playful installation, on to its closure, due to COVID-19, and its subsequent reopening. Our findings explore the challenges caused by both the hybrid nature of the installation, as well as the visitors' playful use of the installation which made the experience social and performative – but also caused some problems. We also discuss the problems \textit{We Dare You} faced in light of hygiene demands due to COVID-19. The analysis contrasts the design processes and expectations of stakeholders with the audience's playful appropriation, which led the stakeholders to see the installation as both a success and a failure. Evaluating the design and redesign through use on behalf of visitors, we argue that an approach that further opens up the post-production experience to a process of continuous redesign based on the user input - what has been termed "design-after-design" – could facilitate the design of similar experiences in the museum and heritage sector, supporting a participatory agenda in the design process, and helping to resolve the tension between stakeholders' expectations and visitors' playful appropriations. 

\end{abstract}



\maketitle

\section{Introduction}
In recent years, the museum sector has increasingly used digital technologies to facilitate immersive experiences - leading the media scholar Jenny Kidd to describe this development as an "immersive turn" \cite{kidd_immersive_2018} in the GLAM sector (galleries, libraries, archives, and museums). Frequently, such experiences rely on Virtual Reality (VR) and similar technologies. However, the use of VR in museums often comes with challenges - according to Hornecker and Ciolfi, "VR has been one of the most problematic technologies to bring into heritage settings in terms of cost, physical and technical setup, and obsolescence" \cite[p. 45]{hornecker2019human}.

In this article, we follow the lifecycle of a Substitutional Reality (SR) \cite{suzuki_substitutional_2012,simeone_substitutional_2015} installation, a type of installation that makes use of Virtual Reality (VR) technology and maps the virtual experience onto the physical environment and physical objects, setting up a combination of visual and tactile stimuli sometimes referred to as "passive haptics" \cite{insko_passive_2001}. SR experiences allow users to touch and feel physical objects, while viewing virtual counterparts of these objects inside the virtual environment through a VR headset. By having the physical and virtual overlap and diverge at critical moments, SR experiences enhance the total user experience, allowing both the physical and virtual elements to support each other.

SR is a promising technology that has been explored in artistic and technological experiments \cite{tennent_thresholds_2020_jocch,spence_vrtefacts_2020_dis}. However, it also introduces some challenges for visitor experience design, where digital experiences and physical installations are often approached in different ways, both technically, organizationally, aesthetically, and with regard to individual GLAM stakeholders' dissemination goals. 

In this article, we explore these challenges through a study of \textit{We Dare You}, an SR installation in the Danish Architecture Center in Copenhagen. \textit{We Dare You} places visitors inside a virtual/physical environment that sets up a sensory illusion of walking on a plank extending out from the building's facade, daring visitors to jump. Experiences from the design and deployment of \textit{We Dare You} offer great potential to explore the challenges involved in designing an SR experience in the GLAM domain. Using an ethnographic approach, we studied the design process, implementation, and day-to-day running of \textit{We Dare You}, in order to analyze the ways in which various key stakeholders and visitors affected the design and use of the installation. Drawing on the empirical data we have gathered, our paper discusses the challenges and design opportunities of designing a SR installation in a GLAM space through the installation’s lifecycle; including a period of lockdown due to the COVID-19 pandemic, which mandated new procedures for safe use in a COVID-19 context.

Our analysis shows the ways in which the combination of virtual and physical stimuli allowed for playful engagement among visitors, yet how this also became a point of contention among stakeholders. We discuss how similar projects could work strategically to meet similar challenges, by opening up the post-production experience to a process of continuous redesign based on the user input through what has been termed "design-after-design" \cite{redstrom_re:definitions_2008}, thus supporting a participatory agenda \cite{simon_participatory_2010}, which is of key importance for contemporary GLAM practices. Such design processes may help GLAM actors discover and benefit from the successful design elements of playful SR installations, while simultaneously alleviating the tension between stakeholders' expectations and visitors' playful appropriation of those installations.

\subsection{\textit{We Dare You}}

“Do you dare to walk the plank?” This is the invitation greeting visitors to the \textit{We Dare You} installation. \textit{We Dare You} offers a short, playful experience and has been extremely popular with visitors since its launch on 20 July 2019.

\begin{figure}[h]
  \centering
  \begin{minipage}[b]{0.33\textwidth}
    \includegraphics[width=\textwidth]{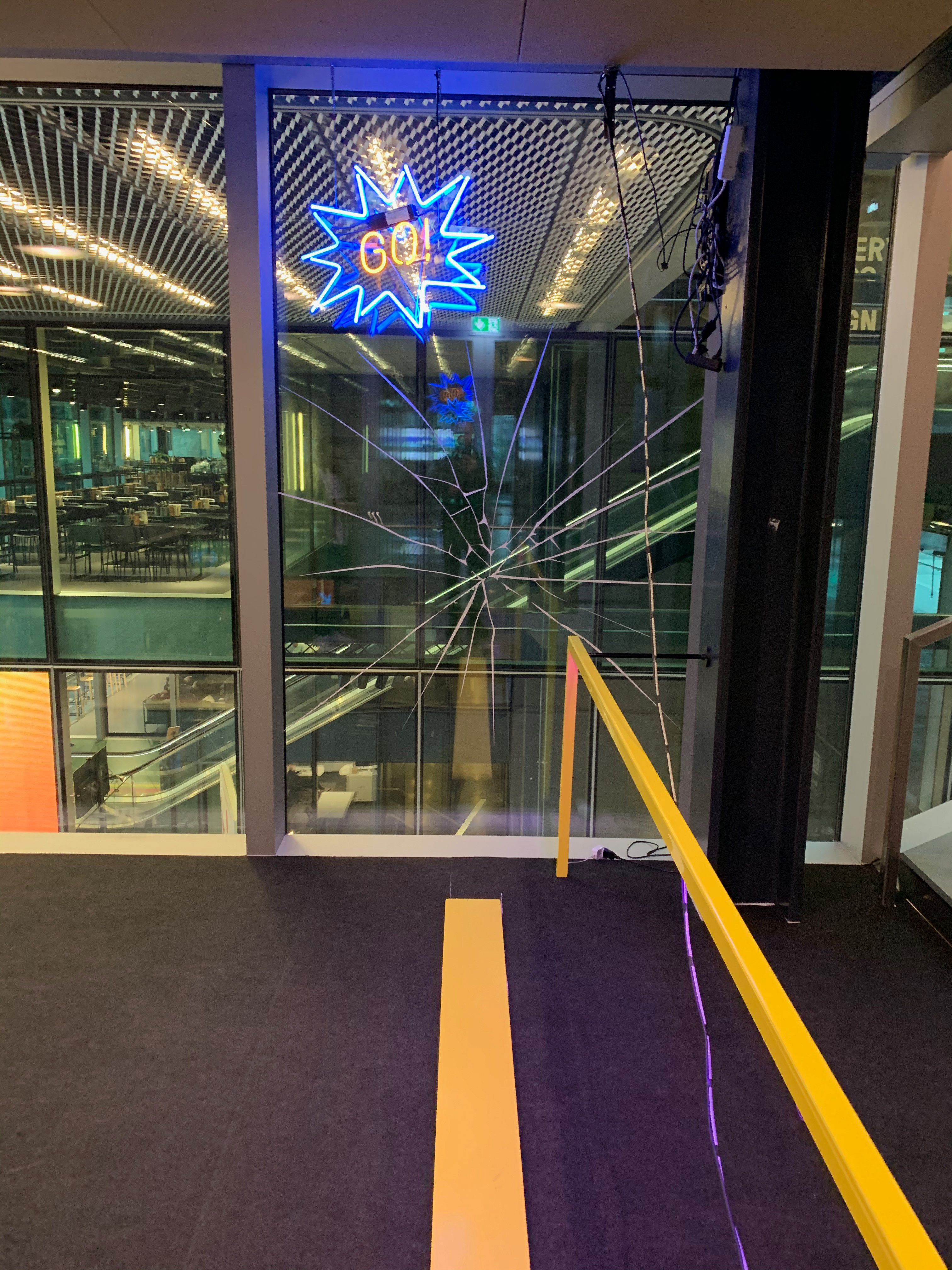}
  \end{minipage}
  \hfill
  \begin{minipage}[b]{0.33\textwidth}
    \includegraphics[width=\textwidth]{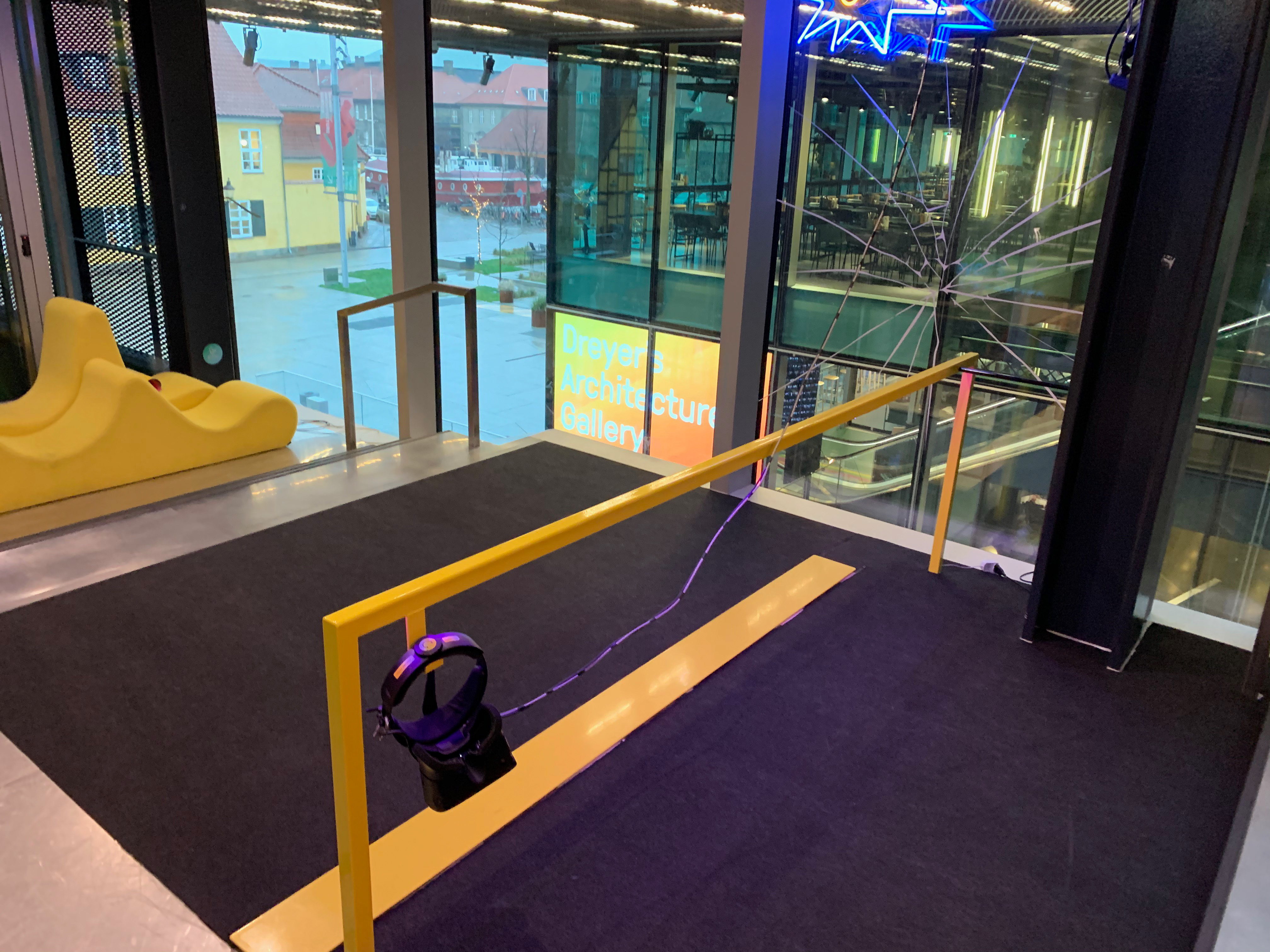}
  \end{minipage}
    \hfill
   \begin{minipage}[b]{0.33\textwidth}
    \includegraphics[width=\textwidth]{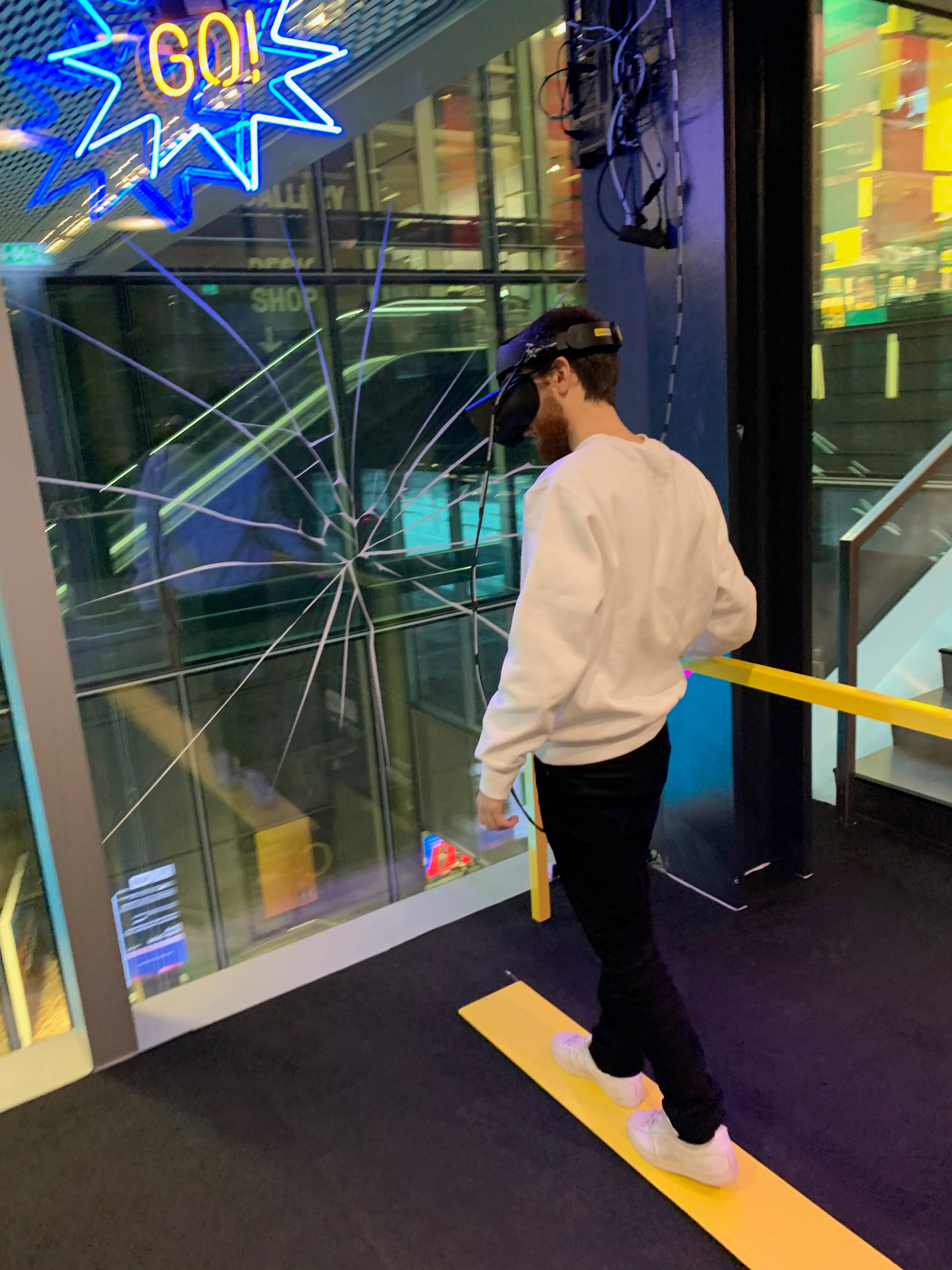}
  \end{minipage}
  \caption{The physical environment of the \textit{We Dare You} experience including handrail and plank.}
  \Description{Image 1, 2, and 3 displaying the \textit{We Dare You} experience including handrail and plank.}
  \label{wedareyou}
\end{figure}

\begin{figure}[h]
  \centering
  \begin{minipage}[b]{0.49\textwidth}
    \includegraphics[width=\textwidth]{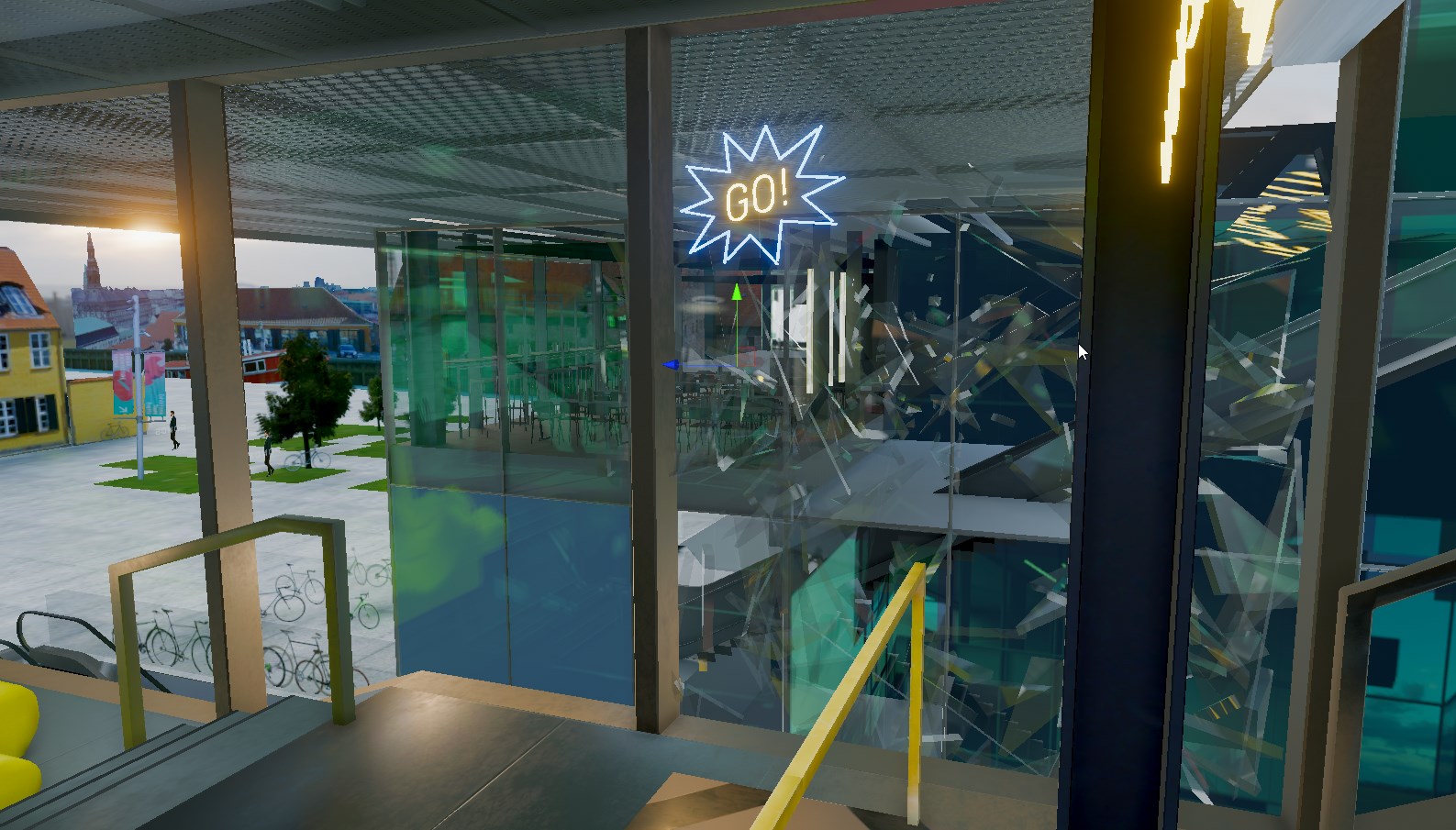}
  \end{minipage}
  \hfill
  \begin{minipage}[b]{0.49\textwidth}
    \includegraphics[width=\textwidth]{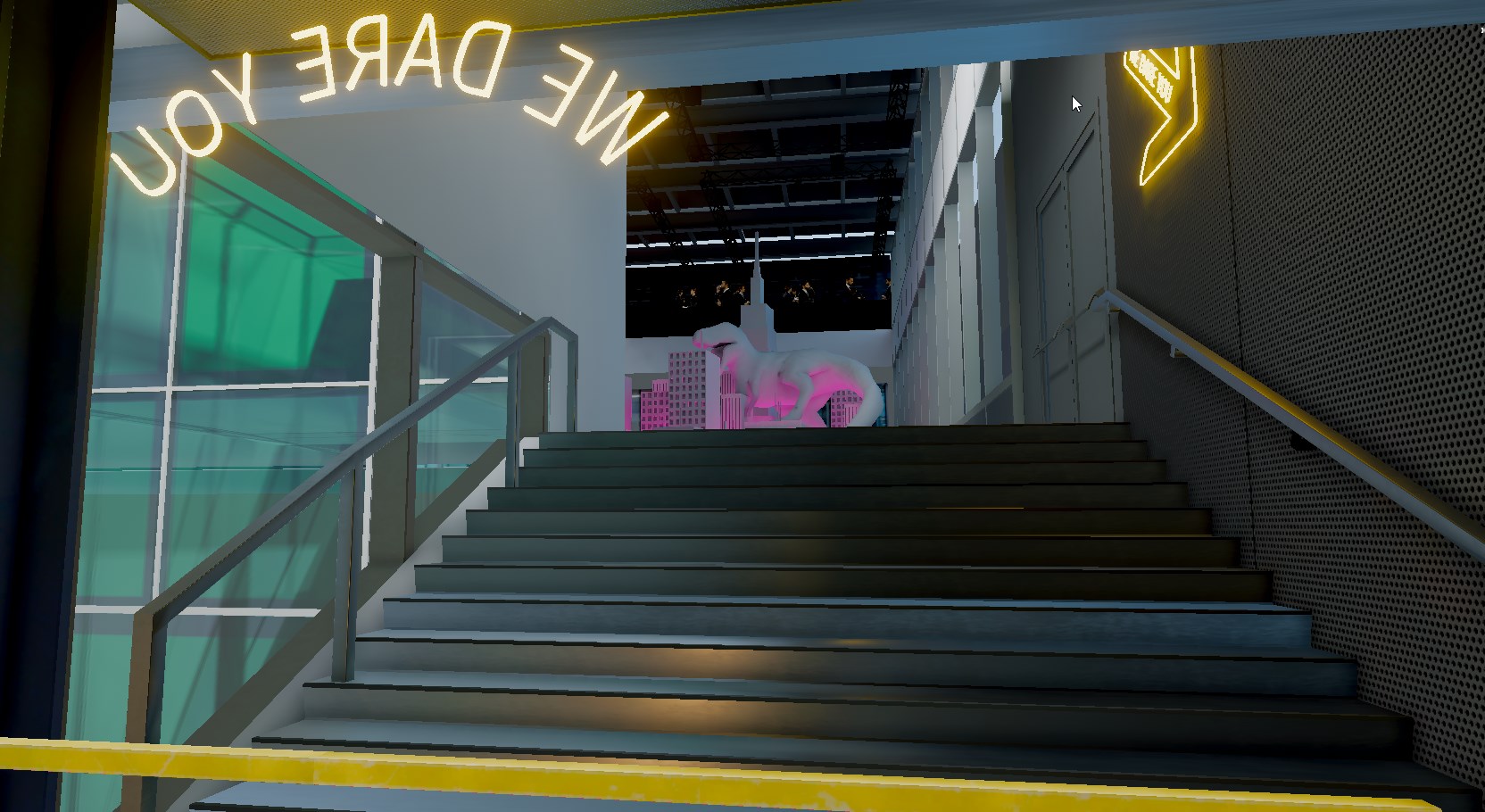}
  \end{minipage}
  \caption{The virtual environment of the\textit{We Dare You} experience.}
  \Description{Image 1 and 2 are displaying the virtual environment of the \textit{We Dare You} experience.}
  \label{virtual_wedareyou}

\end{figure}

Visitors arrive at the installation after passing through the main exhibition space, where temporary exhibits are presented. They encounter a neon sign saying "\textit{We Dare You}" in large glowing letters, in front of a wide staircase that leads down to a yellow metal plank and railing, ushering participants toward a window on the facade of the building (Figure \ref{wedareyou}). A large sticker on the window makes it look like it is cracked. A VR headset hanging on the railing is connected with a long cable to a computer mounted high behind a metal pillar. When visitors put the the headset on, they encounter a space that is a near-perfect replica of the physical space in which they are standing, except that the window appears a bit closer than in real life (Figure \ref{virtual_wedareyou}). This illusion makes it possible for the visitor to walk up to the virtual window, while still remaining at some distance from the physical window in real life. As a visitor approaches the virtual window, the glass cracks, then breaks, and the virtual plank extends outside of the building—extending to match the actual length of the physical plank the visitor is walking on. For participants, this creates the illusion that they are walking out of the window and observing the building from the outside, three floors above the ground. As the visitors reach the end of the plank, they can jump off. This jump is just a short drop in physical space (2 cm), but in the virtual space, it is a long fall to the virtual ground, which breaks as they land, mimicking a “superhero landing.” The participants are free to look around for a moment, and then they are prompted to take the headset off.

\section{Play, participatory engagement, and experience design in Museums}
GLAM institutions have shown an increasing interest in using new technologies to create playful experiences in their spaces \cite{beale_museums_2011, magnenat-thalmann_virtual_2005, bekele_survey_2018, anderson_developing_2010}. Viewed from the perspective of museology, this is part of a broader shift in the museum sector from physical collections towards stories and experiences \cite{vermeeren_museum_2018, drotner_routledge_2018} or from artefact appreciation to a participatory agenda \cite{simon_participatory_2010}. Kidd describes a turn towards “immersive heritage encounters,” including "a ludic turn [...] characterized by increased interest in the application of play and game mechanics" \cite[p.~1]{kidd_immersive_2018}. Opportunities for participation and engagement have become central issues for museums \cite{ciolfi_including_2008}, and digital solutions such as displays, touch interfaces, games, and augmented and virtual reality now proliferate  \cite{beale_museums_2011, witcomb_interactivity_2006}.

Arguably, engagement and participation are key elements in tackling the challenges that modern museums face; such as remaining relevant, modern, democratic, becoming creative spaces, and facilitating a social experience \cite{simon_participatory_2010, hornecker2019human}. Museums are often reliant on attracting visitors in order to become financially viable; offering new technological experiences and play becomes part of that struggle. Visitors can play and engage in the museum and also interact with objects not present in the museum \cite{bannon_hybrid_2005, ciolfi_including_2008,lovlie_gift_2019,magnenat-thalmann_virtual_2005,ryding_monuments_2018,classen2017museum}.  

Even though the concepts of participation, social interaction, creativity, and personalization are all present in the activity of play \cite{sicart_play_2014, caillois_man_2001, bateson2013play, huizinga2008homo}, museums often struggle with incorporating play: games are seen as non-serious and are often perceived to be disconnected from the subject matter of the museum \cite{eklund_lost_2019}. However, research has demonstrated that play, if left open, offers visitors new ways of engaging critically with museum exhibitions, and these attempts can be interpreted as acts of resistance that create new meaning \cite{bergstrom_gaming_2014}. These properties make ludic engagement suitable in a participatory agenda; through playful engagement, users can make the experiences their own. However, play is in its nature hard to control, and tends to resist structure \cite{sicart_play_2014}. Hornecker and Ciolfi \cite{hornecker2019human} identify the loss of authority and control as a point of concern amongst curators. That play can come in conflict with curatorial practices is often seen as a challenge for the design of playful museum experiences (cf. \cite{back_gift_2018}). Indeed, much research has pointed out how hard participatory values in general have been to implement in GLAM institutions, which historically gained their values through a focus on artefacts and collections, rather than offering visitor experiences (e.g. \cite{black_museums_2020,mccall_mmac_2014}). 

\section{Substitutional reality and sensory experiences in museums}                
A key problem with incorporating digital experiences in the GLAM sector is a common concern among museum professionals—that the new digital experience might subsume the physical collection or space, which is the GLAM actor's main attraction and responsibility \cite{back_gift_2018,hsi_study_2003,lyons_designing_2009,petrelli_2013_integrating,walter_museum_1996,wessel_potentials_2007,woodruff_electronic_2001}. Thus, enhancement of existing resources is a central challenge in digital design in these spaces, if GLAM sectors are to keep their physical relevance. Much attention has therefore been given to VR as a technology for visitor experiences in cultural heritage institutions \cite{bekele_survey_2018}. VR and AR have the potential to enhance what GLAM actors already possess and these technologies have, for example, been used to visualize buildings in archaeology \cite{gaucci_reconstructing_2015, giloth_user_2015, liestoel_augmented_2019, secci_virtual_2019} and architecture \cite{liestol_power_2014, liestol_visualization_2014}. One of the advantages of these mixed reality technologies \cite{milgram_taxonomy_1994} is that they offer connections between the physical and virtual.

In recent years, there has been increased interest in the GLAM sector towards experiences that involve more of the senses than just the visual. Losche \cite{losche2006fate} suggests that visceral engagement through the senses is a key element to immersion and authentic experience, which has disappeared from the museum space due to dominance of the visual sense, a sense centered around rationality and detachment. Feldman \cite{feldman2006contact} discusses how the absence of sensory engagement inhibits the unraveling of the sensory complexity of artefacts, prohibiting the understanding of our cultural past, and ultimately rendering museums spaces where cultural memories are subverted. Dudley \cite{dudley2013museum} further calls for a focus on materiality and sensory engagement and criticizes the absence of physical interactions in  the museum space; even when they are present, they tend to be paired with textual information. Furthermore, she comments on how the interesting elements of the materiality of objects are ignored in museum exhibits, along with "the intimate details of people’s physical, sensory – visual, haptic, aural, oral, gustatory, kinaesthetic – engagements with the physical things in question" \cite[p.~6]{dudley2013museum}. Classen \cite{classen2017museum} suggests that attempts to engage with the museum space in a tactile way are to be expected and should be viewed "as meaningful acts of sensory communion with deep cultural roots" \cite[p.~24]{classen2017museum}, and that the spaces within museums and galleries are material environments full of possibilities. Western museums are primarily dominated by the sense of sight, and incorporating other senses "can open up a space for traditional non-Western and women's art forms" \cite[p.~117]{classen2017museum} resulting in a culturally inclusive space \cite{classen2006museum, classen2017museum, dudley2013museum, losche2006fate}. Towards that vision comes sensory museology, suggested by Howes \cite{howes_introduction_2014}, in which sensory experiences are designed to offer visitors an increased aesthetic appreciation of museum exhibits. This turn to incorporate a wider array of the senses in GLAM digital designs is mirrored in Kenderdine, Chan, and Shaw’s concept of embodied museography \cite{kenderdine_pure_2014}. Embodied museography highlights the importance of multisensory qualities, and the authors argue that through embodiment and digital technologies, museums "can create new levels of aesthetic and interpretative experience” \cite[p.~3]{kenderdine_pure_2014}. Neves \cite{neves_multi-sensory_2012}, Randaccio \cite{randaccio_museum_2017}, and Classen \cite{classen2017museum} further discuss how those multisensory qualities create an inclusive museum experience by rendering its space accessible to audiences with disabilities. Wang \cite{wang_museum_2020} states that the engagement of the senses beyond the visual in the museum space has a central role in the creation of immersive experiences, and may positively impact visitor satisfaction and stimulate “emotion, reminiscence, and education“ \cite[p.~16]{wang_museum_2020}.

Studies have explored the question of how to incorporate the senses beyond sight in VR experiences. Skola and Liarokapis \cite{skola_examining_2019} suggest that VR techniques can create the illusion of touch even without any tactile stimuli. Marshall et al. \cite{marshall_sensory_2019} explore the creative potential in misaligning sensory stimuli using VR in conjunction with physical movement in the art installation \textit{VR Playground}, \cite{tennent_twenty_2019} where the user sits on a swing using a VR headset in order to traverse a set of abstract environments. Each environment re-maps the motion of the swing differently, using the kinesthetic qualities of the physical swing as a core part of the experience.

As early as 2001, the concept "passive haptics," in which a visual virtual environment is augmented with physical objects, was suggested to have a strong effect on the user’s sense of presence in a virtual environment \cite{insko_passive_2001}. More recently, passive haptics has been explored by Harley et al. \cite{harley_sensory_2018}, using physical solutions to diegetically engage the senses in VR experiences using the ambient sensory qualities of physical spaces. Chagué and Charbonnier \cite{chague_real_2016} employed, amongst other elements, passive haptics to create a strong feeling of presence in their Real Virtuality platform. Similarly, Campbell et al. \cite{campbell_feeling_2018} use a wheelchair as a tangible user interface, exploiting its passive haptics to create the feeling of physical presence inside a virtual space.

Harley et al. \cite{harley_tangible_2017} have discussed the concept of Tangible VR, where they attempted, through tangible, material objects, to connect the physical with the virtual. Tangible VR is closely connected to another recently introduced concept in the spectrum of mixed reality that attempts to draw on the opportunities of sensory experiences: Substitutional Reality (SR) \cite{simeone_substitutional_2015, suzuki_substitutional_2012}. In SR, physical objects are matched with virtual objects in VR, so that a user can physically touch and feel a physical object while experiencing a digital version of it through a VR headset. In a recent project, Tennent et al. \cite{tennent_thresholds_2020_jocch} used SR to create an art installation with passive haptics in which users could explore a recreation of a museum exhibition from 1839. In the \textit{VRtefacts} project, SR is used to heighten a visitor’s sense of connection and engagement with items from a museum's collection \cite{spence_vrtefacts_2020_dis}. For the museum sector, Substitutional Reality holds potential to recreate inaccessible objects, spaces, and experiences, facilitating sensorial engagement as well as enhancing the existing locale and physical collections of GLAM actors.

\section{Technological appropriation, design-after-design, and ludic engagement}
To create meaningful engagement between the visitor and the museum collection, museums rely on the field of design \cite{black_transforming_2012}. Contemporary GLAM goals often include active and engaged visitors who co-produce their experience, yet these goals have been hard to achieve in practice \cite{drotner_routledge_2018,hornecker2019human,kidd_immersive_2018,simon_participatory_2010,vermeeren_museum_2018}. This problem is mirrored in digital design where a participatory view on users has not always been self-evident. Dunne \cite[p.~71]{dunne_hertzian_2005} argues that when it comes to the design of electronic products, the design field could benefit from considering “the user as a protagonist and co-producer of narrative experiences rather than a passive consumer of a product’s meaning.” That critique is supported by Gaver, who describes playfulness as "an antidote to assumptions that technology should provide clear, efficient solutions to practical problems" \cite[p.~1]{gaver2002designing}. Hornecker and Ciolfi warn that design must take into account the technological appropriation that can emerge in the museum context, and that "designing interactive visitor experiences must allow for a degree of flexibility, and of freedom for visitors to adjust the experience (and related content) to their own evolving interests" \cite[p.~7]{hornecker2019human}.

Redström \cite{redstrom_re:definitions_2008} refers to that appropriation as “design-after-design”: the redesign of an object by its users, as they reformulate and change the meaning and purpose of the object through use. He criticizes dominant practices of user-centered design for relying on user testing of prototypes rather than actual use of the designed object ("use-before-use") — a process that does not account for the appropriation that often occurs through use. Redström suggests, amongst other things, the concept of tactical formlessness as discussed by Hunt \cite{hunt_just_2003}. This means designing things that are not finished, that can change their form after what we traditionally call the design phase of a project is finished. Several scholars have contributed to a re-thinking of the relationship between designed object and user in the design process in order to account for appropriation \cite{ehn_participation_2008, kelly_displacing_2014, van_der_velden_participatory_2014}. Frauenberger \cite{frauenberger_entanglement_2019}, adding to the concept of "design-after-design," argues that our design methods need to adapt to those ideas and cease to distinguish between design and use, and instead treat the “design-use” of things as a process that continually affects the resulting technological artefacts and our relationships to them.
 
 Tactical formlessness and design-after-design promotes dissolution of boundaries which, we argue, is pertinent when trying to support active participation, co-creation, and human activity as amorphous as play. In other words, we can acknowledge that what we design is not finished once the product leaves the hands of the designer. This is an idea well explored in studies of digital games. Games, it is argued, come to be as they are played; thus the player is part of making a game what it is, and sometimes games are changed fundamentally by players playing “wrong” and adding on to the game \cite{consalvo_cheating_2009, prax_co-creative_2012}. In this study, we interpret the design-after-design approach as broadening the design space to also include the decisions that happen after the traditionally named design phase in a project. Concretely, this implies looking at specific aspects, namely use, implementation, and maintenance as part of design. Use, in this sense, is seen as co-producing the experience. By dissolving some of the boundaries in the design process - i.e. the strict division of stages as design, implementation, and maintenance - we can acknowledge how user appropriation is a kind of design (see \cite{bjogvinsson_design_2012}). We argue that such an approach may support a participatory agenda in the GLAM sector. Furthermore, it creates an environment for learning that is based on constructivistic principles \cite[cf.]{hein_learning_1998}. Finally, this approach has the potential to address the emergence of appropriation that comes with play.
 
\section{Method}

For this qualitative study, we employed specific ethnographic methods \cite{hammersley_ethnography_2007}. During this study, the first author has been working as an industrial researcher in the Danish Architecture Center (DAC) as part of his PhD programme, starting four months before the opening of the installation and ongoing. As an employee, he has had privileged access to documentation and could conduct interviews with staff and observations of the installation. We draw on knowledge gained from being immersed in the Center’s work, participating in informal conversations with the DAC's staff, as well as formal, structured interviews with five key project stakeholders and two members of the floor staff~(\ref{subsection:interviews}) and observations of the experience~(\ref{subsection:observations}). However, it should be noted that the first author was not involved in the commissioning, design, or maintenance of the \textit{We Dare You} installation. The role of an industrial PhD gives him an independent role as researcher, allowing him to view the project from a critical distance. The other two authors are university researchers with no direct involvement with DAC.

    \subsection{Interviews}
    \label{subsection:interviews}
    Three months after the installation was set up (October 2019), we interviewed five key stakeholders from the \textit{We Dare You} project in order to evaluate the results of the whole process of implementation. We interviewed two project managers, the head of program, the designer, and the IT support staff. We conducted the interviews via audio recording during working hours at the DAC, and they were subsequently transcribed. In order to thoroughly compare different perspectives of the various stakeholders, we crafted a set interview guide for structured interviews \cite{berg_qualitative_2008} where as much relevant information as possible could be gained in the limited time slots set aside for our interviews, due to the schedules of the stakeholders. A set of introductory questions mapped the informants’ background with VR and the project itself. We wanted to see through which lenses each participant would view the installation. Understanding their experience with Virtual Reality, and the specifics of their involvement with the project, allowed us to understand their values, desires, concerns, and motives. In addition, it served as a means to facilitate the construction of a narrative by the person being interviewed. Through that narrative, they were able to describe their experience with the \textit{We Dare You} project. In a second set of questions, the goal was to explore the successes and challenges of the collaboration and project cycle. The questions were designed to gain information about the specific aspects that interested us: the roles of the various stakeholders, including the designer, the role of the DAC, and reflections on the experience. It is important to note here that even though we refer to this installation as a substitutional reality installation—as explained above—the stakeholders refer to it as Virtual Reality or VR. Two more interviews were conducted with two of the “hosts”—floor staff responsible for interacting with the visitors, both to provide help and answer questions, and to have stimulating and critical discussions about architecture. Those interviews aimed to get direct information and insights about what occurs in the exhibition space on a daily basis. In the presentation of our findings we draw on quotes from the interviews, and all informants were made aware of this and gave their permission. A few participants wanted the opportunity to read through their own quotes before publication. In order to protect the informants, we use pseudonyms instead of their real names.
    
    The interview data was analyzed using a content analysis \cite{berg_qualitative_2008} approach, which combined deductive and inductive elements. Thus, themes were drawn out based on our engagement with the data as well as previous literature. In the inductive process, we identified several themes relating to the work with \textit{We Dare You}. The interview results were compared with insights from observations. In particular, we were interested in instances where the DAC's staff and the designer's opinions and impressions seemed to differ from what visitors did with the installation. Thus, we searched the data for key points of disagreement and disruption, points where tensions could be identified and explored.

    \subsection{Observations and documentary data}
    \label{subsection:observations}

    Informal observations of the installation happened regularly where the first author observed both visitor and staff interactions with the exhibition. From 11 to 13 February 2020, the first author spent three hours per day conducting formal observations with systematic note-taking, observing approximately 150 visitors using the installation in total. For two out of those three days, the first author was doing the observations while operating as a volunteer floor staff, standing by the installation and assisting with any technical issues that occurred. This option was preferred as a less intrusive way of gathering observational data. The observations were written down as they were observed during the three-hour shift and later analyzed in conjunction with the results of the interviews.

The DAC also gave us access to data from a set of standard questionnaires that the floor staff routinely filled out every day, wherein they report their impressions and feedback from visitors as well as technical and practical issues that emerges. These data cover the time period from 20 July 2019 until 12 January 2020, and provided us with insights about how the installation was experienced by visitors during that time (as seen through the observations of staff).

\section{The Lifecycle of We Dare You}

In this section, we are presenting a rich description of the lifecycle of the installation, connecting together insights from interviews, observations, and floor staff questionnaires.

\subsection{Conception}
\textit{We Dare You} was the result of an open call for submissions for a new VR exhibit at the DAC Center.

\begin{quote}
We always try to find new ways of making representations of architecture more interesting. So, VR is interesting for me as a project leader and also for DAC as a new way to explore how to communicate and create spatial experiences for our guests. (Robin, Project Manager)
\end{quote}

The Center was interested in creating something new for their visitors, something which went beyond their regular exhibits. They wanted to show the building from a new perspective that was inaccessible from the physical space. To achieve this, they announced an open call for collaborators and settled on a proposal from the \textit{Immersive Stories studio} \cite{immersivestories} and the \textit{Khora studio} \cite{khora} to collaboratively develop the \textit{We Dare You} experience. After the software development was finished, a team from the DAC created the physical environment where the \textit{We Dare You} installation was placed, in dialogue with the designer. It was set up in July 2019 and has since been located on the third floor in the exhibition space, in front of a glass wall that overlooks the entrance to the center (Figure \ref{wedareyou}).

The experience has a physical and a virtual part, and is accessed through VR Goggles (Oculus Rift S). The virtual elements were developed in the Unity environment \cite{unitywebsite}. Similar experiences have been developed in the past, such as Richie’s Plank Experience \cite{plankwebsite}, an experience that allows the player to experience walking off the top of a skyscraper. In this experience, players use their own physical plank laid out in their home, while a virtual plank inside the virtual game environment is automatically adjusted to fit the physical plank. A key difference between these two experiences is that Richie's Plank Experience takes place in a fictional space that has no connection with the physical space of the user (other than the plank), while the virtual space of \textit{We Dare You} is an accurate recreation of the physical space the visitor is standing in. Using photogrammetry, the physical space of the DAC was mapped in order to be replicated in its virtual counterpart. That process was done prior to the installation of \textit{We Dare You}, therefore photogrammetry was used to capture the building's facade and exhibition space without the physical elements of the installation. The designer and the DAC thus took different responsibilities in the project, and executed their designs at different times, with the designer first creating the virtual environment and the DAC then building the physical interaction elements. Those physical elements initially consisted primarily of the signage material, plank, window sticker, and a hand-railing for the visitors to hold onto to make sure they would be able to walk on the plank and not fall off while wearing the headset. 

As stated in the quote from the project manager above, the DAC expected the installation to allow visitors to engage with architecture in new, interesting ways. This was also the designer's intention:

\begin{quote}
[We Dare You] reveals the important elements of what architecture does to you because it triggers your senses in a way that you feel the anxiety when you enter the virtual outdoor room. It shows how the senses are triggered and what a physical room does to you. And it also reveals to you perspectives of the architecture of [the DAC building] since you are not able otherwise to stay at this tunnel outside the window. And then on the fun side, it aims at a youth audience as its target group. (Simone, Designer)
\end{quote}

The stakeholders saw the SR technology as attractive for visitors due to the novelty of the technology. That attraction was especially appealing to the type of visitors they wanted to attract: families and younger audiences. The project was seen as helping the institution to offer a cutting-edge experience, which would make the DAC seem up-to-date and relevant.

\begin{quote}
It’s interesting because there is a nice energy in the VR environment. People are excited, trying out new things, I really feel it is some sort of frontier so that kind of energy was nice (Robin, Project Manager).
\end{quote}

During development and as the project led up to launch, a sense of excitement and doing something truly innovative characterized the project. The DAC's staff saw the experience as an opportunity to engage visitors—in particular younger audiences—in new ways; to give them a “fun” experience in contrast to the more serious character of the rest of the exhibition. All stakeholders connected the installation with words like fun, interactivity, and experience.

\subsection{The \textit{We Dare You} Launch: Audience Hit, Technological Failure}
Once the experience launched, the installation was highly successful in terms of visitor satisfaction. Staff reported that visitors enjoyed it, were excited to try it out, and that some of them visited the DAC specifically to try \textit{We Dare You}—this is mentioned frequently in the visitor feedback given to the DAC staff as reported in the floor staff questionnaires. Often, long lines of visitors waiting to try the experience were observed. However, during the first three months, the exhibit frequently had problems with software and hardware failure. First, the VR headset required frequent re-calibration. The floor of the exhibition space is slightly reflective, and combined with the fact that the natural changes in the light throughout the day, caused the sensors to require re-calibration. This issue was addressed by placing a non-reflective carpet on the floor of \textit{We Dare You}, significantly reducing this problem. The experience still requires re-calibration occasionally, but significantly less often than before. A second issue arose from the way the software was configured to trigger the start of the experience, which did not adequately take into account the behaviors of users. The software was configured to start when a new player put on the headset, and then stepped in front of the plank. However, many visitors would first step on the plank, then put the headset on. Unfortunately, the software was not adapted to this situation, as the signs prompting visitors to move into the starting position would now appear behind them in the virtual space. Thus, these visitors would walk on the plank without the virtual experience running, leaving them either on the VR idle screen, or on the second part of the experience (after the super-hero landing). This software behavior was a source of confusion, causing the visitors to walk back, trying to find the beginning of the experience, and occasionally, when that did not work, they would abandon the installation altogether. To address this issue, DAC placed a sticker on the floor, indicating where visitors needed to stand when they put the headset on to start the experience (see picture 4). A more flexible software solution would have been difficult to implement, since DAC does not have access to the Unity source code, nor the expertise to edit it. Therefore, the software remains the same since the release day.

A number of issues also occurred relating to the Oculus headset and the cable connecting it to the stationary computer. \textit{We Dare You} requires the visitors to walk forward, jump, take off the headset, turn, and walk backwards. Since people would typically turn towards the same direction, the cable tended to twist and eventually break. Sometimes visitors would also move too far from the computer that the headset was connected to, causing it to disconnect. Both of these issues were addressed by supporting the cable with a metal cable that prevented excessive twisting and kept the cable in place. The headset was also frequently dropped and damaged, either by accident when visitors would take it off their heads, or due to children running with the headset and jumping at the edge of the plank, hitting their face and the headset on the window. At the point of writing, four headsets have needed to be replaced due to the resulting damages. Due to this problem, a sticker was placed on the floor, prompting parents to not leave their children unattended, and to take care of the VR equipment. Since then, the headset and the cable suffer less damage, however, it is uncertain if that is due to that sticker that acts as a reminder, or if it is due to the metal cable making the installation more robust. 

Most of these solutions were implemented over the first three months. During this time, constant IT intervention was required. The solutions greatly reduced the attention the installation required, but even at the time of writing, \textit{We Dare You} requires occasional IT intervention due to malfunctions.

\begin{quote}
With We Dare You, I think it has probably been available to visitors less than 50\% of the time. I think if you look at the records from the exhibition hosts, they are calling facilities at least 3-4 times a week, maybe more (Chris, Project Manager).
\end{quote}

Around half of all the comments in the floor staff questionnaires report technical errors or visitors being frustrated about not being able to try the installation when it was out of order.  In their feedback to the hosts, visitors reported disappointment or anger that they were not informed in advance that the installation was not working on the day of their visit. 

\begin{figure}[h]
  \centering
  \begin{minipage}[b]{0.49\textwidth}
    \includegraphics[width=\textwidth]{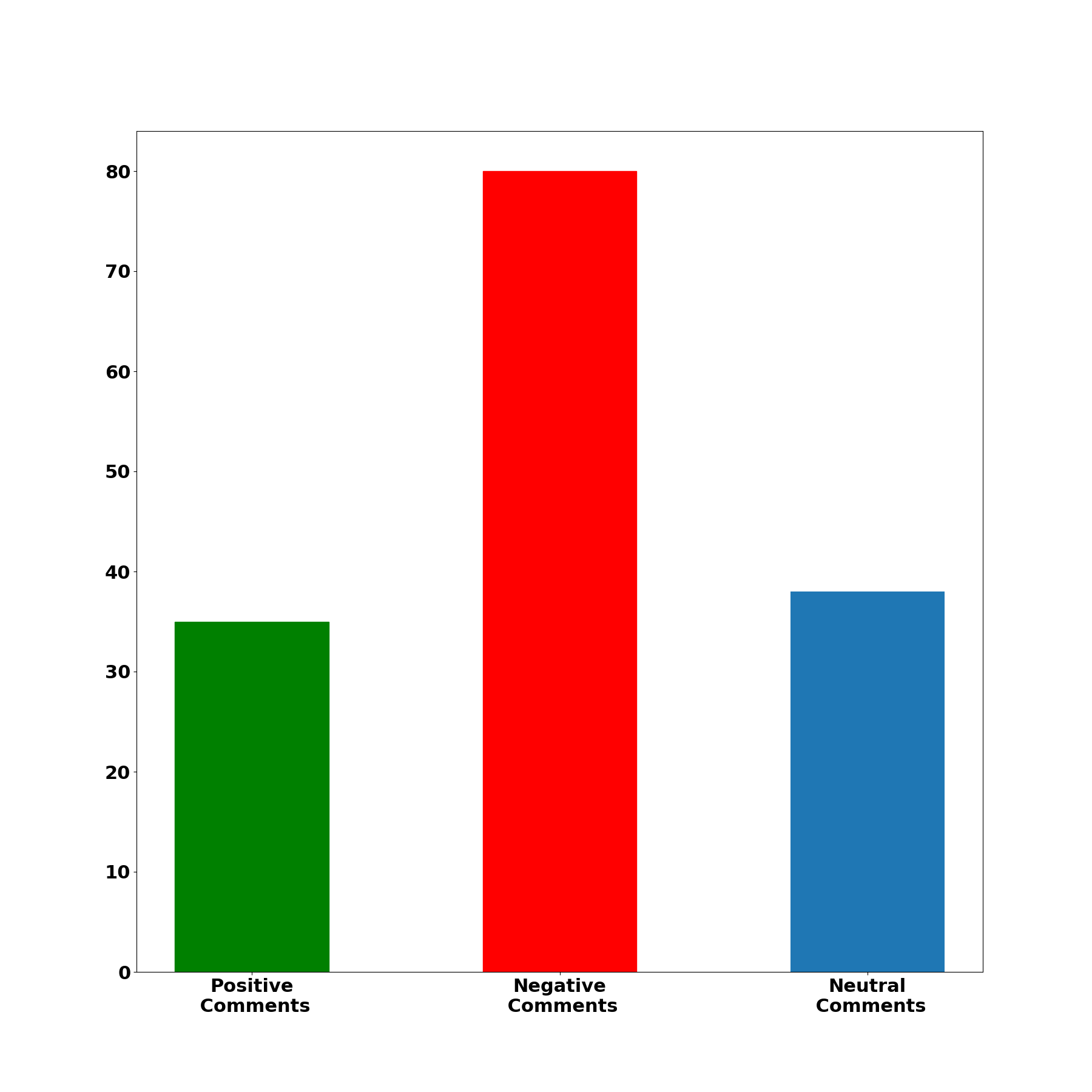}
  \end{minipage}
  \hfill
  \caption{Deductive content analysis on comments in floor staff questionnaire responses that mention the installation (N=153).}
  \Description{}
  \label{q1}
\end{figure}

Some of the nature and extent of these problems can be captured by looking at the data from floor staff questionnaires, covering the first six months after the launch of the installation. These questionnaires are filled out by the center's "hosts," which are an important part of the center's offer to visitors. Their role is not only to assist the visitors and answer their questions, but also to engage in critical discussion about architecture. Furthermore, they are responsible for fixing minor technical issues that may arise in the exhibition space, and to inform facilities of more serious technical difficulties. After their shifts, hosts answer a set of questions that describe what occurred during their shift. In that data set, the installation is mentioned numerous times. Figure \ref{q1} displays the amount of comments that mention the “\textit{We Dare You}” installation: In the 276 questionnaires filled out in the time period, there were 101 questionnaires mentioning \textit{We Dare You}, with 153 comments in total (some questionnaires included more than one comment about the installation). The large amount of comments related to \textit{We Dare You} clearly demonstrates that the installation has had an important impact on the visitor experience, from the perspective of the hosts. Furthermore, the figure also shows the results of a deductive content analysis separating the comments into positive, negative, and neutral comments.

\begin{figure}[h]
  \begin{minipage}[b]{0.49\textwidth}
    \includegraphics[width=\textwidth]{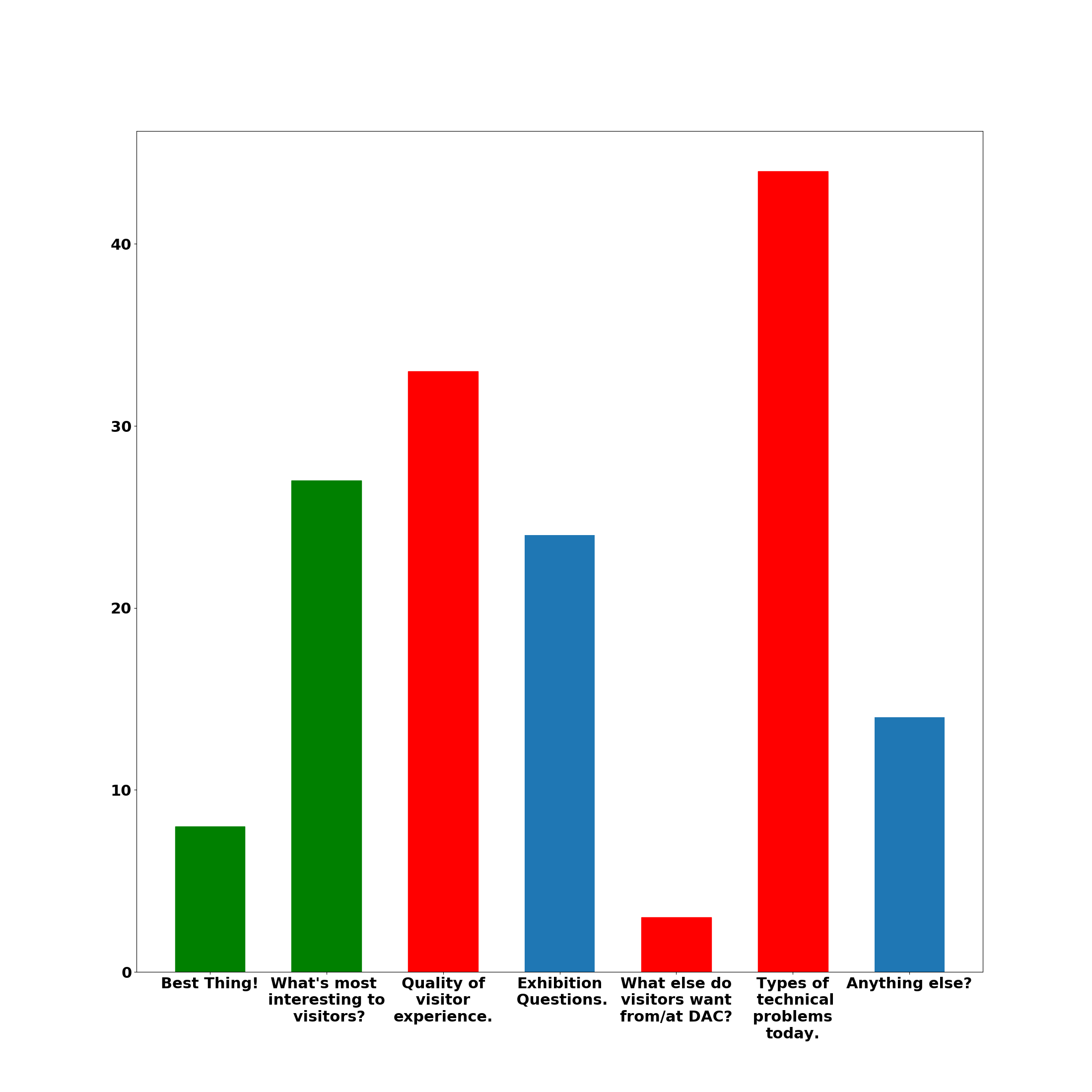}
  \end{minipage}
    \hfill
  \caption{Number of floor staff responses per questionnaire category that mention the installation (N=153).}
  \Description{}
  \label{q2}
\end{figure}

Figure \ref{q2} separates the \textit{We Dare You}-related comments into categories corresponding to different questions in the questionnaires. The color of the columns indicate the sentiment of the majority of comments in the relevant category: green for positive, red for negative, and blue for neutral sentiment. The vast majority of the comments that are related to problems — represented as red in the figures — are related to technical issues with the installation, rather than other aspects of the experience the visitors had with the installation. In other words, even though there are many negative responses to the question about "Quality of visitor experience" relating to \textit{We Dare You}, these comments generally refer to technical problems with the installation. It is worth noting that \textit{We Dare You} was mentioned many times under the category "What's most interesting to visitors?", as well as a few times under "Best Thing!" (see figure \ref{q2})—reflecting the generally positive response from visitors when the installation was not affected by technical issues.

\begin{quote}
Then, I think that for our visitors, VR is not new, and they are quite familiar with putting on glasses or using different kinds of digital equipment, and that also means that they are quite rough on the equipment. ... today children—and also grown-ups but especially young people and children—are so used to them, so for them it is like grabbing any other kind of instrument. That creates some new demands: hardware and the user robustness. That’s something that we have to solve (Terry, Head of Programme).
\end{quote}

 DAC’s preconceived notions of careful use of equipment did not match up to the reality of what people did with the installation, nor with the attitudes and experiences people had. Once launched, visitors used the full installation or just elements of it in various playful ways that diverged from the predicted use. For example, we observed children hanging from the handrail without the headset on or running on the plank with the headset on, jumping at the end of it, and on rare occasions even hitting the physical window. The social and performative elements of watching others engage with the experience have rendered it successful with groups of visitors such as families or friends visiting together. As can be read in the quote above, visitors, and in particular children, do not treat the VR objects with the same reverence that DAC perhaps expected. Instead of having serious, profound experiences with architecture, play reshaped the meaning of \textit{We Dare You} from technologically ground-breaking and offering visitors new experiences with architecture to playful appropriation.

\begin{quote}
I think the evaluation is quite clear, we need to figure out some more robust hardware that can actually take all the people coming in, wanting to explore (Terry, Head of Programme).
\end{quote}

As in the quote above, some of our stakeholders saw what visitors were doing as something to which the institution could adapt. Others raised concerns as they were trying to make sense of the new experience in relation to DAC's mission to educate—and not just entertain—visitors about architecture.

\subsection{Success and Tensions with Play}
\begin{figure}[h]
  \centering
  \includegraphics[width=0.7\linewidth]{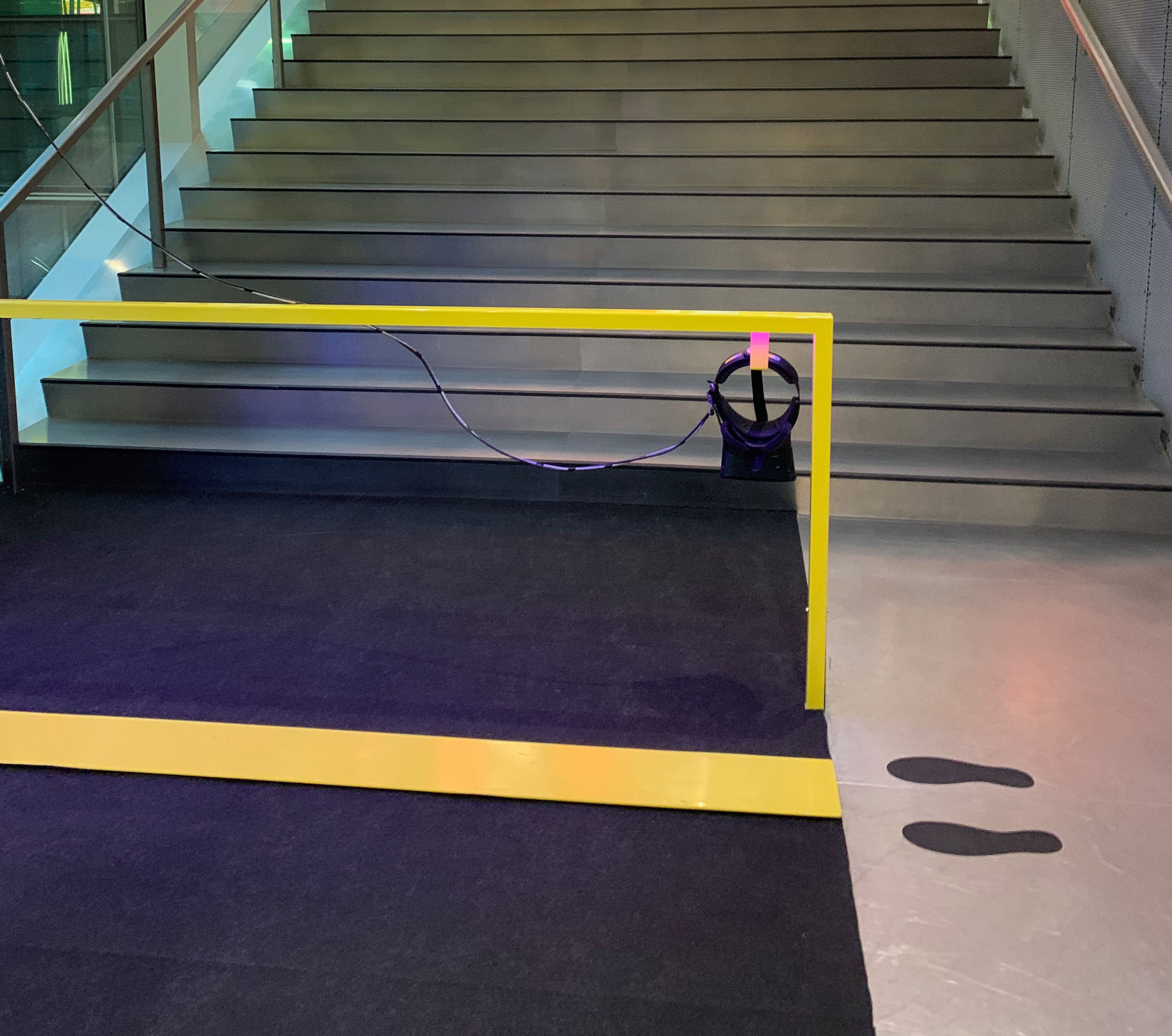}
  \caption{Stairs next to the installation, where visitors wait in line and watch the users of the installation.}
    \Description{Stairs next to the installation, where visitors wait in line and watch the users of the installation.}
      \label{stairs}
    \end{figure}

After the initial period of constant breakdowns and subsequent redesign of the physical elements, DAC managed to increase the robustness of the installation. It was clear that \textit{We Dare You} was a great success according to visitors and it did succeed in attracting a new, younger demographic to the DAC, evident in both interviews and observations. We commonly observed crowds during peak hours sitting on the staircase located next to the installation (figure \ref{stairs}). That staircase was appropriated by visitors in two distinct ways. They used it to queue when waiting to try the installation, or used it as a resting point after having been through the main exhibition space, which offered them an entertaining view of the people using the installation. It was also common—especially for children—to try the installation multiple times, so the stairs served as a point where parents relaxed and watched their children play with the installation.

\begin{quote}
We still could have had that playful moment of the jump, but I would have liked to see the steps before that, of how do we help people understand the architecture, and that I think is missing right now… If we had asked the question of "how do we want people to better understand our building, the design of our building, and use the technology to go and see those places and those views that I cannot see other ways" then we still could have had the last step being "jump off the building" (Chris, Project Manager).
\end{quote}

As the technical problems were solved, other concerns remained. The playfulness which made the exhibit a success was a contested point among the DAC staff. As in the quote above, play became positioned as the opposite of a meaningful experience with architecture. As visitors interacted with the exhibit, they did not do so in a way which favored serious engagement, but rather playful exploration.

\begin{quote}
I don’t think they [visitors] get that much out of it architecturally, to be honest. It is a way to experience architecture of course and people also look around as they have fallen and it is a way to experience architecture but I don’t know if people are that aware of it. It is a way to be aware of the building, but mostly it is a fun experience (Oliv, Floor Host).
\end{quote}

Play came to be contrasted with properly experiencing the architecture. The informant above expresses that most visitors simply see the fun in the experience, and play rather than engage with it in order to have more serious experiences. A playful exhibit cannot be controlled or contained in the same way as a non-playful one. As soon as DAC invited visitors to engage and to play, they lost control over how visitors behaved. These playful experiences did not match up the expectations the stakeholders had.

\subsection{New Perspectives: An Embodied and Social Experience}

The sensory engagement afforded by substitutional reality offered a new quality to the visitor experience. \textit{We Dare You} was designed to invoke the senses and create a feeling of vertigo when walking the plank. Many visitors had intensely embodied experiences with the \textit{We Dare You} installation. A host explained that many visitors using the experience held hands with a friend or family member to gain courage to engage in the experience and combat the fear of heights, which sometimes stopped people from engaging.

\begin{quote}
You know that you are here [in the exhibition] but it tricks you! It’s too scary! (Comment by visitor during the observations 11th of February 2020)         
\end{quote}

Contrary to many conventional exhibitions that rely primarily on objects and text and thus focus on stimulating the mind, \textit{We Dare You} engages the visitor’s body as part of the visit. This embodiment holds the potential of making the experience extraordinary and memorable. 

\begin{quote}
VR is a way for us to engage the bodies of our guests and that’s a good thing because the body remembers and the brain remembers. (Robin, Project Manager)
\end{quote}

While unforeseen and not planned for, once installed \textit{We Dare You} became a social experience. Due to the physical elements of the plank and handrail (see figure 1), it was easy for visitors to see what was going on, the digital experience became externalized through the physical elements. As visitors engaged, others could sit on the staircase and observe, and the built environment made that observation interesting. The absence of the virtual environment revealed the illusion for observers, which made the visitor’s behavior such as stepping nervously, hesitating before jumping, and so on, seem comical.

\begin{quote}
[It’s] more immersive and fun [...] because it’s not many people who want to be overloaded with information about the material and the construction and all the other technical drawing or thinking behind a building like this. (Terry, Head of Programme) 
\end{quote}

Due to its playful and performative nature, the installation serves as a location in the exhibition where visitors choose to film each other: parents film their children, friends and couples film one another. We attribute this behavior to the aspects of performance afforded by the physical elements of the installation. Those physical elements allow visitors to act out their responses to the virtual stimuli, in such a way that is entertaining for the surrounding audience. This is not unique behavior but it gave the installation a further performative dimension as users acted out for the camera: an artifactual status that comes from the elements of physical interaction—plank, handrail—and from the behavior of the user of the installation having elements of performance—being brave, being scared, screaming, laughing, et cetera.

\begin{quote}
The nice thing about the installation is that it’s fun to see people wearing the goggles because you can see they are scared and they don’t know whether they want to jump or not. So it is exactly with this experience and the exhibition design, that makes it fun to watch. The user becomes part of the installation. (Robin, Project Manager)    
\end{quote}

The physical setup of the experience created a spectacle for on-lookers, drew attention, and created a social atmosphere for visitors waiting or just passing by. The staircase next to the installation, in which visitors would line up to try the experience, gave the effect of a small spectator stand, setting the space of the installation up as a scene for a performance. This setup helped solve a common concern with experiences that use VR headsets in GLAM spaces: the isolation of the visitors from their companions, since what is in the headset can only be experienced by the one wearing it. The \textit{We Dare You} installation demonstrated that substitutional reality has the potential to afford stronger social experiences than typical VR experiences.

\begin{quote}
I am not worried about the technologies or the experiences themselves, I am more worried about how do we design the physical structures around it, how does it become a social activity, or how does it become fun for people who are lost in the VR goggles. We always bash VR experiences for being private experiences that are boring for onlookers, but in the case of We Dare You, it is fun for onlookers. As an architectural museum, how can we take it further than the goggles so that the physical environment is also interesting? (Robin, Project Manager) 
\end{quote}

While many VR experiences have problems with virtual motion sickness, no such issues are present in our data. However, we do not know if that is because the visitors did not experience motion sickness or because it was too mild to report on, since the experience was so brief.

\subsection{\textit{We Dare You} and COVID-19}
On the 13th of March 2020, the installation—along with the rest of the DAC—was shut down temporarily following orders issued by the Danish government in response to the COVID-19 pandemic. On the 1st of June 2020, the DAC re-opened, and \textit{We Dare You} opened shortly after. In order to follow new hygiene regulations and public demands on measures to combat further spread of the virus, DAC placed hand sanitizer and disinfectant wipes next to the installation. Additionally, a government-issued sign reminds visitors to keep their distance and use sanitizer frequently. 

The DAC has also changed their cleaning procedures and now cleans the entire facility once per day. \textit{We Dare You} is currently open and in operation. At the time of writing (November 2020), technical issues such as the cable disconnecting, or the sensors requiring re-calibration still arise on a weekly basis; however, the processes set in place by the institution are more efficient, addressing those issues quicker than they used to, partly due to practical know-how built up from past experiences.

\begin{figure}[h]
  \centering
  \begin{minipage}[b]{0.33\textwidth}
    \includegraphics[width=\textwidth, angle=270]{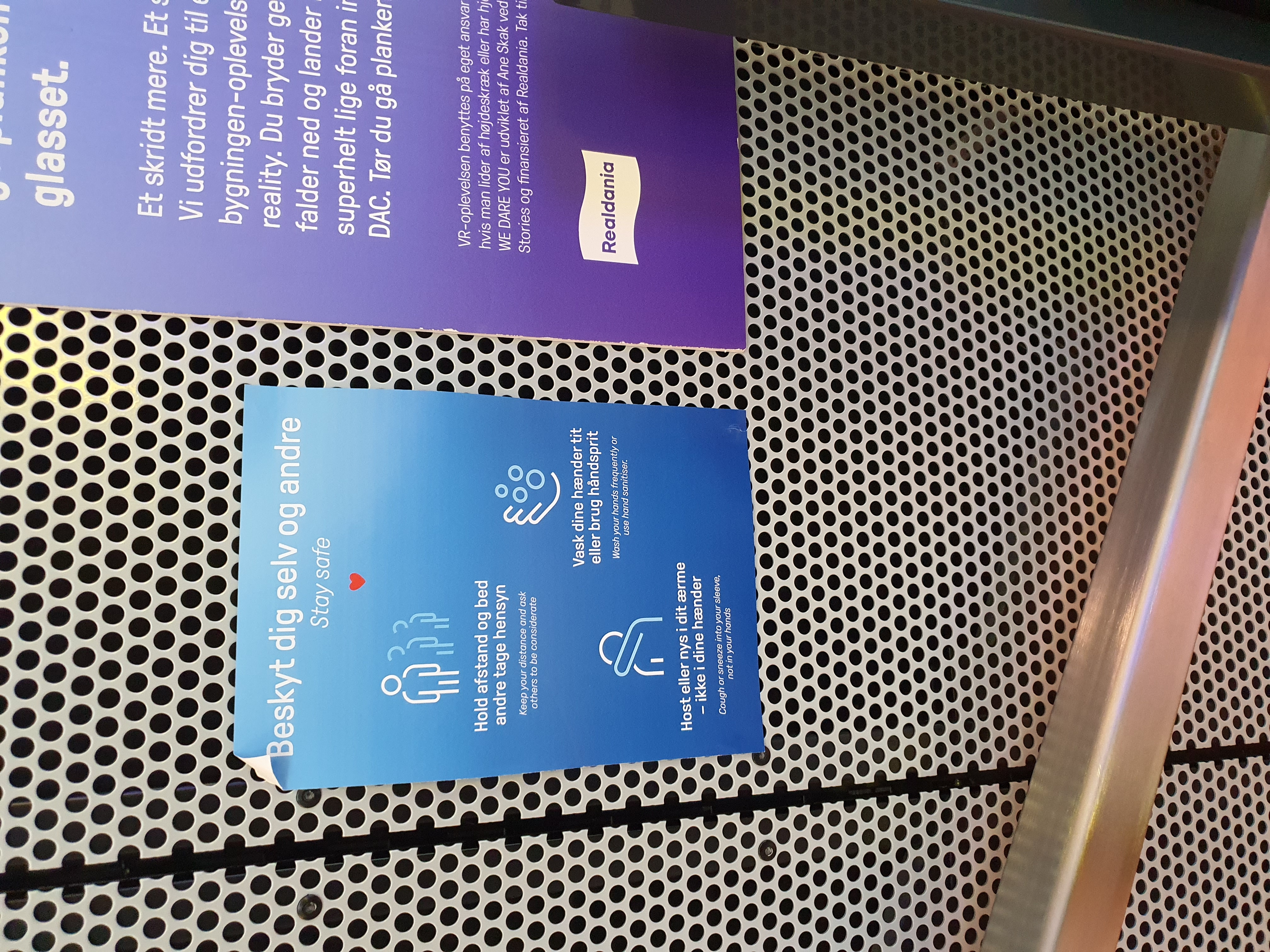}
  \end{minipage}
  \hfill
  \begin{minipage}[b]{0.33\textwidth}
    \includegraphics[width=\textwidth, angle=270]{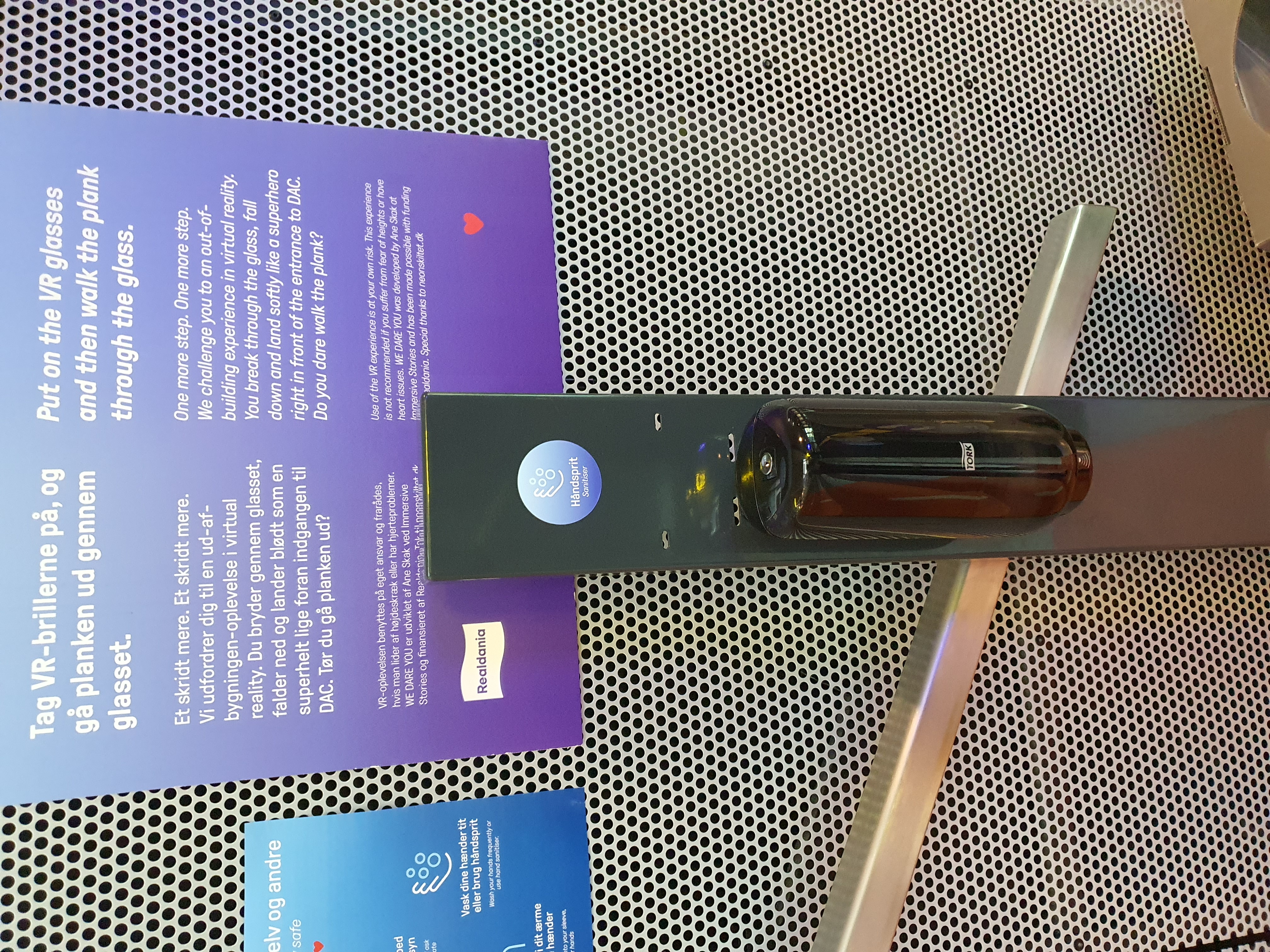}
  \end{minipage}
    \hfill
   \begin{minipage}[b]{0.33\textwidth}
    \includegraphics[width=\textwidth, angle=270]{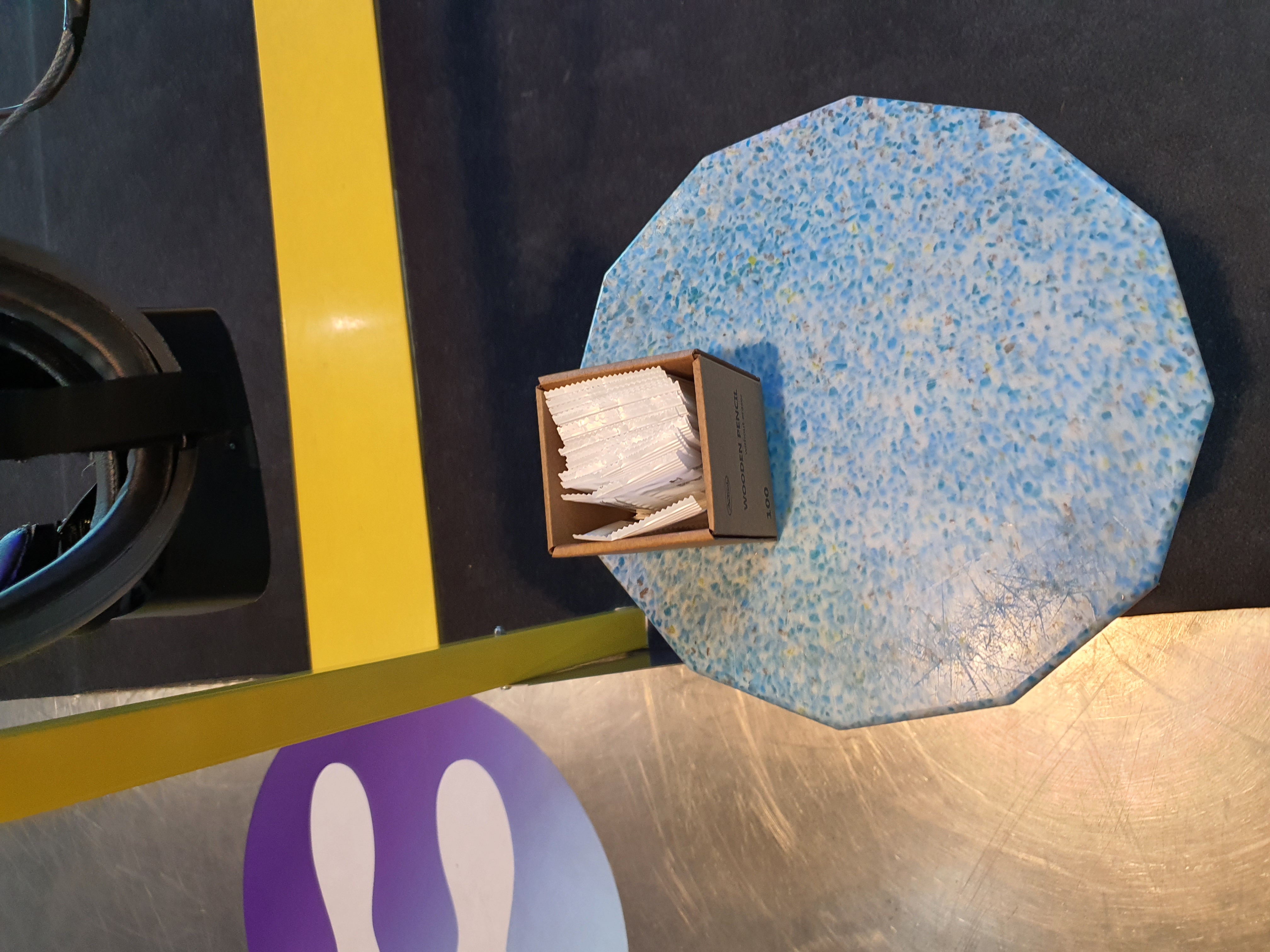}
  \end{minipage}
  \caption{COVID-19 related information and sanitizing equipment}
  \label{wedareyou_corona}
\end{figure}

\section{Discussion}

While the \textit{We Dare You} installation was a success among visitors, it also brought a number of challenges for the DAC. First, the fragility of the technological setup caused much downtime and a need for constant repairs. The installation also required several adjustments to the setup. These problems were exacerbated as the contract released the designer from all responsibilities to help maintain and update the installation after launch. The reason for this unfortunate arrangement can probably be found in the DAC's regular procurement process, which normally contracts physical exhibits that are easier to maintain, and for which the center has significant in-house expertise. Such challenges are likely to arise in GLAMs using similar project-based approaches, if those procedures do not include plans for maintenance and continuous involvement from the designer after the launch.

Second, visitors used the exhibit in ways the stakeholders had not foreseen, playfully exploring it and engaging in performances that challenged DAC's intentions with the exhibit. Visitors playfully explored the installation: running, jumping high, walking backwards, walking outside the plank, and so on. While these creative uses may have contributed to the technical problems, they were also the source of joyful engagement for visitors and created space for play, performativity, social interaction, and spectacle. However, some of the stakeholders and floor staff were concerned that the entertainment value of the experience had come at the expense of DAC's core mission: that visitors should learn about architecture. The ludic turn \cite{kidd_immersive_2018} in cultural institutions creates demands that professionals engage with play in settings often valued on providing education and dealing with serious topics. Play, while often being positioned in opposition to seriousness and rationality \cite{grimes2009}, has strong transformative potentials. Play can transform how we look upon and define activities going on in GLAMs. This goes some way towards explaining the uneasy relationship with play we see at the DAC.

One might imagine a number of strategies for handling such challenges in future projects of a similar kind. First, the issues relating to technical fragility and maintenance seem to call for an improved procurement process tailored to the needs of interactive installations, especially those that invite physical play with fragile technology. Such a process might include a more robust, iterative design process with rigorous testing and approval before launch, a use-before-use approach focused on user testing, and problem solving before launch. However, there is no guarantee that such as process would catch all the problems; in the case of \textit{We Dare You}, most of the issues arose from the exhibition environment and the unexpected behaviors of museum visitors, which would be difficult to fully replicate in a testing situation prior to launch. This is a common problem when it comes to interactive museum installations \cite{hornecker2019human}. Second, the concerns regarding education might similarly be addressed by incorporating such concerns in the design process, for instance by specifying the learning requirements as part of the design brief, and developing procedures for evaluating the educational outcomes of the installation. Such an approach would have the advantage that it would force the organization to make explicit success criteria for the educational dimension of interactive exhibits, which might help designers and museum staff explore the trade-off between learning and other qualities such as engagement, playfulness, and curiosity. However, an increased focus on education would risk coming at the expense of the fun and engaging qualities of the installation, a focus that also cannot be enforced in "free-choice learning" spaces like museums \cite{hornecker2019human}, where visitors follow their own interests.

Instead, we suggest that it might be more productive to reframe some of the issues described here, seeing them not only as problems but also as a productive set of tensions that can be used to create interactive experiences that offer visitors more freedom to play and explore, while also facilitating contemporary forms of learning. This could be achieved by adopting a design-after-design approach, by perceiving the appropriation from visitors as a reinterpretation of the museum experience—design through use. Such an approach could be employed by GLAM actors to fulfill ideals of visitor participation \cite{simon_participatory_2010} and support a constructivist pedagogical view \cite{hein_learning_1998,hornecker2019human} where visitors are encouraged to make their own meaning, while supporting the fun and engagement that emerges through play.

Redström \cite{redstrom_re:definitions_2008} describes “Design after Design” as an approach that entails designing artefacts without a clearly prescribed use, creating a space of possibility that leaves it open to users to define the use of the object. Following a design-after-design approach in the process of conceptualizing, developing, and deploying installations can allow GLAM actors to explore what happens as visitors use their exhibits as opportunities for participation and appropriation. This opens up concrete strategies for co-designing and co-creation through including visitors in shaping their visit. Design-after-design can act as a catalyst helping GLAM actors to provide interesting experiences for visitors, by opening up a dialog between the institution and its visitors that alters and guides the design of an installation.

Such an approach entails designing interactive installations that are open for a wide variety of uses. The institution can then evaluate those interactions and their effects on the institution's dissemination goals. Using the results of that evaluation, the institution can adapt the design to accommodate the interactions that the visitors choose to engage with, while continuing to support the institution's dissemination goals. This process shares similarities to a traditional prototyping cycle, but it differs in the fact that it maintains a constant dialogue, even after the deployment of the installation. That constant dialogue between visitors and GLAM actors affects the role of an installation in the actors' spaces. We can learn from the changes visitors make, and instead of trying to resolve the contrast to the imagined use, look upon it as part of the design.

A design-after-design inspired development would entail that future exhibits make explicit a participatory agenda and this is already an active policy goal for many GLAM institutions. This entails accepting that the institution in question does not have full control over the experience they are offering visitors, nor can they predetermine exactly what visitors will take away from the experience. However, as the \textit{We Dare You} example shows, this may often be the case anyway. Using this as a guiding principle from the start, much of the tension arising between various goals and expectations by designers, GLAM professionals, and visitors could be resolved through dialogue and awareness of the active part visitors play in shaping the final experience.

Designing installations that are open to appropriation can be achieved in several ways. Some key guiding principles are flexibility, openness, and configurability \cite{ehn_participation_2008}. Through such a design-after-design perspective it is possible to change the expectations of the exhibit's outcomes while the use of the exhibit re-negotiates its intended meaning. Key is empowering users to engage in meaning-making themselves, with the GLAM providing an open framework, rather than a fixed experience.

A design-after-design approach has the potential to support visitor agency through their participation in the act of design-through-use. Established use-before-use approaches fall short in accommodating the needs of a GLAM context, from the multitude of target groups that are present, to the dynamic environment that changes during various exhibitions of different characters. The type of design thinking suggested by design-after-design results in a co-creation process for installations, establishing the role of GLAM sites as democratic and creative spaces that can provide interesting experiences that adapt to the ever-changing exhibition environment and can be enjoyed by different visitor target groups, while supporting the GLAM actors' individual dissemination goals. We aim to explore the challenges and potentials of this endeavour through an ongoing research-through-design \cite{zimmerman_research_2007} project.
 
\section{Conclusion}
By all accounts, the \textit{We Dare You} installation has been a great success with the DAC’s visitors—aside from the frustrations with the technical problems. However, as shown above, DAC stakeholders had somewhat diverging views of the installation: some see problems mainly of a practical nature (e.g. a need for more robust hardware, more testing, and resources for maintenance), whereas some take issue with the lack of educational content in the experience and the dominance of play. This illustrates that the GLAM stakeholders, depending on their role in the organization, have diverging—and perhaps even conflicting—demands for the installation. To some, it is most important that it attracts audiences (and, in particular, young people), to others it is more important that it supports learning, whereas to yet another group it may be more relevant to see this as an experiment showcasing the organization’s research and innovation efforts. In our experience, these conflicting views are quite typical of efforts to set up playful technologies in museums \cite{back_gift_2018, waern_hybrid_nodate}, and highlights the complex demands placed on GLAM actors by funding bodies and society at large regarding project budget, engagement of broad audience groups, dissemination of knowledge, research, marketing, and more.

While one might argue that these demands make GLAMs a challenging domain for playful installations in general, the case at hand brings out some considerations of particular importance when designing for Substitutional Reality. First of all, the fact that SR is a hybrid technology with both physical and virtual components means that the design process may need to combine two different design disciplines — on the one hand, interaction design (and software development), and on the other hand, product design/exhibition design with physical materials. The first area includes a wide variety of concerns familiar to HCI scholars, such as the need to set up agile/iterative design and development processes, including users in design and testing, and to design for unexpected behaviors and appropriation by users. The second area includes a number of concerns and expertise from curators and other GLAM sector professionals, such as the connection to the physical space, the robustness of materials, physical safety rules, universal access, and more. DAC and other GLAM institutions have much experience and expertise with the latter area, but much not with the former; and in the \textit{We Dare You} project, DAC chose to delegate the responsibility for the design of the virtual components to an external designer, whereas the organization itself took charge of designing the physical components. This division of labor, common for GLAM actors, may have complicated the process. As an implication for design, designers of SR installations should be wary of design processes that decouple responsibility for the physical part of the installation from the virtual, and instead plan design processes that provide for close coordination in the design of both physical and virtual components. Furthermore, the responsibility for the design should not stop at the launch of the installation, but rather allow for continuous work to adapt the installation to the emerging behaviors of users and corresponding challenges that arise from this - as suggested by the design-after-design perspective.

Second, the \textit{We Dare You} installation demonstrates a promising solution to a common problem with VR installations: That of providing an interesting experience to people who are waiting for their turn. Often, this is done through screens that display what the user can see with the VR headset, offering the onlookers a window into the VR experience. \textit{We Dare You} does not do this. Instead, the SR's close mapping between the physical and virtual space, as well as the fact that the user is interacting with the virtual space by moving through the physical space, means that it is possible for onlookers to directly observe the user’s movements through the installation. The discrepancy between physical and virtual space—leading the user to feel like they are balancing on a plank high above the ground outside the window—while in the physical space, the plank is just two cm above the floor — makes the situation comic. As noted above, this has been a great source of entertainment for the visitors, and has contributed to the popularity of the installation. A similar effect was achieved by the creators of Thresholds \cite{tennent_thresholds_2020_jocch} where those standing in line could look into the exhibition space through a window in the wall, whereas the participants wearing VR headsets would, instead of the window, see a particularly eye-catching painting, often leading them to come close and lean towards the window, giving the onlookers a comical view into the SR space. This demonstrates that SR offers additional design opportunities compared to VR: As GLAM visits are usually social experiences, it is an important (albeit often overlooked) requirement to design experiences that can be shared. VR is challenging in this regard, due to the reliance on headsets that block interaction with those outside the virtual space. The physicality of SR thus offers opportunities that might be valuable both in the GLAM context, as well as in other contexts where the social experience is important.

The closure and subsequent reopening of the \textit{We Dare You} installation due to the COVID-19 crisis also points to a challenge for SR and VR that is pressing at this point in time, but may also have far-reaching consequences. The current pandemic has posed a whole new set of criteria for sustainable experience designs in the GLAM sector, and elsewhere. In a recent piece in the Smithsonian magazine \cite{billock_how_2020}, Jennifer Billock maps out how public life and architecture changed in New York after the war on tuberculosis in the late nineteen hundreds. Billock uses this analysis to ask what changes museums and other public spaces will need to implement in the years to come to combat the ongoing and future pandemics. Hygiene concerns, situations with queues and buildup of large gatherings of people are key problems going forward. These issues have particular implications for SR and sensory museum experiences as the pandemic has "forced a de-prioritization of touch and physicality" \cite[p. 298]{galani_hybrid_2020}. A key challenge is: how can GLAM institutions deploy and maintain playful SR installations or indeed any installation relying on VR headsets, or similar technologies? Two issues merit particular attention. First, to accommodate for the constant flow of visitors, it must be possible to sterilize devices quickly and easily. At DAC, this is currently solved by giving the visitors the responsibility for disinfecting the equipment before and after use. Whether this will be acceptable in the long term remains to be seen. Second, the social element of play poses a design challenge, since it often has the effect of gathering people closer together through playful interaction, sometimes amongst strangers, or through audience formation. Anecdotal evidence from observing the exhibit since its reopening suggests that visitors often forget to hold the mandated distance while looking at and engaging with the installation. These are issues that need to be addressed so that installations like this can find stable places inside exhibition spaces in the foreseeable future. While this is an urgent need in the current crisis, it is likely that the pandemic may lead to a heightened awareness towards the risk of contagious diseases in general for many years going forward, making contagion-safe interaction a long-lasting requirement for experience design in public spaces such as GLAMs. It is an important task for HCI to document cases such as this, in order to provide education for future design projects.

\begin{acks}
This study is funded by the Innovationsfonden and Realdania. Furthermore,
it is part of the GIFT Project, which has received funding from the European Union’s Horizon 2020 research and innovation programme under grant agreement No 727040.
\end{acks}

\bibliographystyle{ACM-Reference-Format}
\bibliography{bib2}

\end{document}